\documentclass[usenatbib]{mn2e}
\usepackage{natbib}
\usepackage{amssymb}
\usepackage{epsfig}
\usepackage{rotating} 
\usepackage{aas_macros}

\newcommand{\Msol}{\ensuremath{\mathrm{M_{\odot}}}}

\newcommand{\rt}{\ensuremath{R_{\mathrm{200}}}}

\newcommand{\Zsol}{\ensuremath{\mathrm{Z_{\odot}}}}

\newcommand{\egc}{{\it e.g.}}  
\newcommand{\etal}{{\it et al.\thinspace}}

\newcommand{\Chandra}{\emph{Chandra}\ }

\newcommand{\ROSAT}{\emph{ROSAT}\ }
\newcommand{\XMM}{\emph{XMM-Newton}\ }

\newcommand{\MEKAL}{\textsc{MeKaL}\ }

\newcommand{\chisq}{\ensuremath{\chi^2}}

\newcommand{\gta}{\,\rlap{\raise 0.5ex\hbox{$>$}}{\lower 1.0ex\hbox{$\sim$}}\,}  
\newcommand{\lta}{\,\rlap{\raise 0.5ex\hbox{$<$}}{\lower 1.0ex\hbox{$\sim$}}\,}  

\newcommand{\nm}{\mbox{\ensuremath{\mathrm{~\nm}}}}
\newcommand{\cm}{\mbox{\ensuremath{\mathrm{~cm}}}}
\newcommand{\m}{\mbox{\ensuremath{\mathrm{~m}}}}
\newcommand{\km}{\mbox{\ensuremath{\mathrm{~km}}}}

\newcommand{\kpc}{\mbox{\ensuremath{\mathrm{~kpc}}}}
\newcommand{\Mpc}{\mbox{\ensuremath{\mathrm{~Mpc}}}}
\newcommand{\s}{\mbox{\ensuremath{\mathrm{~s}}}}
\newcommand{\ks}{\mbox{\ensuremath{\mathrm{~ks}}}}

\newcommand{\keV}{\mbox{\ensuremath{\mathrm{~keV}}}}
\newcommand{\erg}{\mbox{\ensuremath{\mathrm{~erg}}}}

\newcommand{\arcm}{\ensuremath{\mathrm{^\prime}}}
\newcommand{\arcs}{\arcm\hskip -0.1em\arcm}

\newcommand{\mags}{\mbox{\ensuremath{\mathrm{~mag}}}}

\newcommand{\cmsq}{\ensuremath{\mathrm{\cm^2}}}

\newcommand{\pcmsq}{\mbox{\ensuremath{\mathrm{~cm^{-2}}}}}
\newcommand{\pMpc}{\ensuremath{\mathrm{\Mpc^{-1}}}}
\newcommand{\ps}{\ensuremath{\mathrm{\s^{-1}}}}

\newcommand{\ctpp}{\ensuremath{\mathrm{ ~counts ~pixel^{-1}}}}

\newcommand{\ergps}{\ensuremath{\mathrm{\erg \ps}}}
\newcommand{\flux}{\ensuremath{\mathrm{\erg \ps \pcmsq}}}
\newcommand{\ent}{\ensuremath{\mathrm{\keV \cmsq}}}

\newcommand{\kmpspMpc}{\ensuremath{\mathrm{\km \ps \pMpc\,}}}

\newcommand{\LCDM}{$\Lambda$CDM~}

\begin{document}


\title[\XMM observations of ClJ0046.3$+$8530]{\XMM observations of the relaxed, high-redshift galaxy cluster ClJ0046.3$+$8530 at $z=0.62$}
\author[B.J. Maughan \etal]
  {B. J. Maughan,$^1$\thanks{E-mail: bjm@star.sr.bham.ac.uk}
    L. R. Jones,$^1$ D. Lumb,$^2$ H. Ebeling,$^3$ and P. Gondoin$^2$\\ 
  $^1$School of Physics and Astronomy, The University of Birmingham,
  Edgbaston, Birmingham B15 2TT, UK\\
  $^2$Science Payloads Technology Divn, Research and Science Support Dept. of ESA, ESTEC, 2200 AG Noordwijk, The Netherlands\\
  $^3$Institute for Astronomy, 2680 Woodlawn Drive, Honolulu, HI 96822, USA\\ 
}

\maketitle

\begin{abstract}
A detailed analysis of \XMM observations of ClJ0046.3$+$8530
($z=0.624$) is presented. The cluster has a moderate temperature
($kT=4.1\pm0.3\keV$) and appears to be relaxed. Emission is detected
at $\ge3\sigma$ significance to a radius of $88\%$ of $\rt$ (the radius enclosing an overdensity of 200 times the critical density at $z=0.624$) in a
surface-brightness profile. A temperature profile (extending to
$0.7\rt$), and hardness-ratio map show no significant departures
from the global temperature, within the statistical limits of the data. The cluster's
bolometric X-ray luminosity is $L_X=(4.3\pm0.3)\times10^{44}\ergps$, and the
total mass derived within $\rt$, assuming isothermality and hydrostatic equilibrium, is $M_{200}=3.0^{+0.6}_{-0.5}\times10^{14}\Msol$. The gas entropy measured at $0.1\rt$ is compared with a sample of local systems, and found to be consistent with self-similar evolution with redshift. The metallicity, gas density profile slope, and gas mass fraction are all consistent with those found in low-z clusters.
\end{abstract}


\section{Introduction} \label{sect:intro}
The X-ray properties of galaxy clusters, particularly at high redshift, are very sensitive probes of cosmology \citep[\egc][]{hen97,jon98a,arn02a,all03}. Cluster mass is the property most directly related to the predictions of cosmological models, but also one of the most difficult to measure. X-ray observations directly measure the properties of the gaseous intra-cluster medium (ICM), which can be used to derive the total gravitating mass of a cluster, provided that the gas is in hydrostatic equilibrium. This derivation also requires knowledge of the temperature structure of the gas. Thus to measure accurate cluster masses with X-ray observations, detailed measurements of the dynamical state of the ICM and its temperature structure are needed. Such observations of local clusters are becoming common, but detailed studies, and particularly measurements of ICM temperature structure at high z ($z\ge0.6$), are very rare \citep{jel01,arn02b,mau04a}.

ClJ0046.3$+$8530 was discovered as an extended X-ray source in the Wide Angle \ROSAT Pointed Survey \citep[WARPS:][]{sch97,per02}, and confirmed as a galaxy cluster with redshift $z=0.624$. The identification was based on its extended X-ray nature, an overdensity of faint galaxies, and concordant optical spectroscopic redshifts of two galaxies, including the the galaxy coincident with the X-ray centroid \citep{hor04}. The redshift of the cluster is confirmed below via an X-ray redshift measurement. Prior to the observation detailed here, relatively little was known about the cluster's properties. ClJ0046.3$+$8530 falls $11\arcm$ off-axis in two consecutive \XMM observations of the open star cluster NGC 188, taken on the $29-30$ October 2000 (observation identifiers 0100640101 and 0100640201) using the medium filter. The analysis of these data are described below, and the properties of the system are derived and discussed.

A \LCDM cosmology of $H_0=70\kmpspMpc$, and $\Omega_M=0.3$ ($\Omega_\Lambda=0.7$) is adopted throughout, and all errors are quoted at the $68\%$ level. At the cluster's redshift, $1\arcs$ corresponds to $6.8\kpc$ in this cosmology. The virial radius is defined here as $\rt$, the radius within which the mean density is $200$ times the critical density at the redshift of observation.

\section{Optical Observations}\label{sect:optical}
The first optical follow-up observation of ClJ0046.3+8530 was obtained
in August 1997 in the form of a $3\times 4$ minute exposure in the R band
with the University of Hawaii's 2.2m telescope. Mediocre atmospheric
conditions combined with a high airmass of 2.2 resulted in poor seeing
of 1.2 arcsec as measured off the combined final image. Deeper
imaging was performed subsequently in August 2001 in the R and I bands (40 and 20 min
respectively) with the 4.2m William-Herschel Telescope (WHT). Again, the
seeing was 1.2 arcsec. All the imaging
data were reduced in the standard manner. Observations of 6 standard
stars of \citet{lan92} were used to calibrate the WHT photometry.
Galaxy number counts made in an area excluding the cluster 
were consistent with those of \citet{kum01} and \citet{met91}, confirming the accuracy
of the photometry, and indicated that
the limiting magnitude was $R=22$.

Spectroscopic observations of galaxies in the field of ClJ0046.3+8530
were performed on the Keck 10m telescope using the LRIS low-dispersion
spectrograph \citep{oke95} in August 1997. The instrumental setup
combined a 2048$\times$2048 Tektronix CCD with 24$\mu$m pixels
(LRIS-R CCD2), the 300/5000 grating, the GG495 order blocking filter,
and a longslit of $1.5\arcs$ width to obtain a dispersion of
$2.5$ \AA/pixel over a spectral range extending from 5150 to $8950\AA$. Two
galaxies were observed for a total of 900 seconds. The airmass of
the targets was 2.4, the seeing 1.0 arcsec. Standard image processing
techniques were used to reduce the obtained data, and the measured redshifts are given in Table \ref{tab:pos}.

\section{Data Preparation} \label{sect:dataprep}
The two \XMM observations of ClJ0046.3$+$8530 were filtered, cleaned and prepared in the standard way (see \citet{mau04a} for details), to give two events lists for each detector. Both observations were affected by brief periods of high background, and their removal left $30\ks$ (observation 1) and $50\ks$ (observation 2) of good time. After cleaning, the mean PN count rates in the $10-15\keV$ energy band (excluding bright sources) were consistent with that found in the \citet{lum02} blank-sky background files, indicating that the high-background periods were successfully removed.

The position of ClJ0046.3$+$8530 near the edge of the field of view raised important calibration issues. The $90\%$ encircled energy radius of the Point Spread Function (PSF) at $1.5\keV$ increases from $47\arcs$ on-axis to $52\arcs$ at the cluster centroid for the MOS detectors, and from $51\arcs$ to $65\arcs$ for the PN detector. These values were derived with the \emph{Calview} tool in the \XMM Science Analysis Software (SAS). Following recent advice from the \XMM help-desk\footnote{See http://xmm.vilspa.esa.es/xmmhelp/Calibration?id=9356 for a discussion of PSF issues.} these encircled energies were computed using the `EXTENDED' accuracy setting in \emph{Calview}. This is the most accurate method for estimating encircled energies, and is valid at all off-axis angles. These encircled energy radii are therefore more up-to-date than the values given in the \XMM user's handbook v$2.1$. In addition, the shape of the PSF changes from being radially symmetric at the optical axis, to a more arc-like shape, elongated orthogonally to the radial direction. In our modelling of the cluster's surface-brightness distribution, where the shape of the PSF is important, model PSFs are computed with the `MEDIUM' accuracy setting in \emph{Calview}, which provides the only two-dimensional models of the PSF. 

\begin{figure}
\begin{center}
\scalebox{0.6}{\includegraphics*{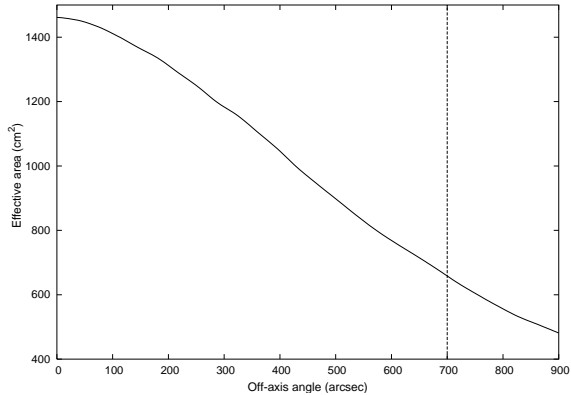}} \\
\caption{\label{fig:effA}The radial dependence of the PN effective area at $1.5\keV$. The dashed line marks the cluster position.}
\end{center}
\end{figure}

The effective area also decreases strongly with off-axis angle. Fig. \ref{fig:effA} shows the dependence of effective area on off-axis angle; at $700\arcs$, the off-axis angle of
the cluster position, the effective area has fallen to less than
50\% of the on-axis value. These instrumental issues are accounted for as much as possible in the following analysis, and their possible effects are discussed.

In addition, the energy-dependent vignetting function of the telescopes has recently
been verified by observations of compact supernova remnants at a range of off-axis angles and azimuths \citep{lum04a}. This analysis revealed a misalignment of $\approx1\arcm$ in the optical
axis of the telescopes containing the PN and MOS2 cameras. For the
cluster location in this observation, the possible discrepancy in the
assumed vignetting would mean that the PN flux has been overestimated by
$\approx6\%$ and the MOS2 flux underestimated by $\approx8\%$. 
A secondary effect of the energy dependence would be that the corresponding temperatures would also be slightly over- and under-estimated respectively. These effects are not accounted for in the following analysis, as the uncertainties introduced are small compared to the statistical uncertainties.

\section{X-ray imaging analysis}
\begin{figure}
\begin{center}
\scalebox{0.45}{\includegraphics*{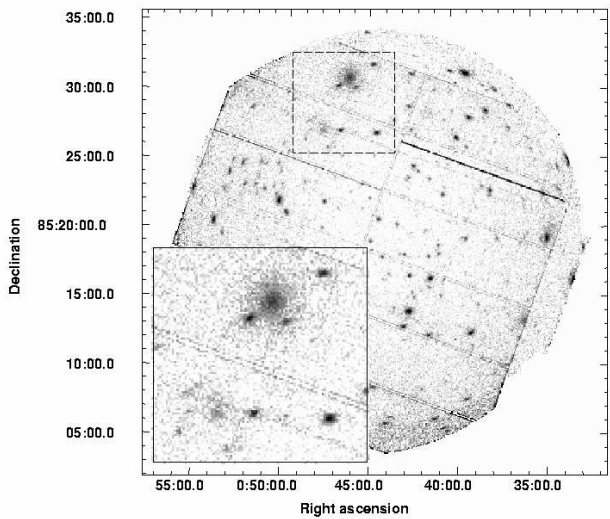}} \\
\caption{\label{fig:fov}An exposure-corrected image of the \XMM field of view. Data from both observations are combined, and binned with image pixels of $4.4\arcs$. The inset is a blow-up of the dashed region, and clearly shows ClJ0046.3$+$8530 and XMMU J004751.7$+$852722 as extended sources.}
\end{center}
\end{figure}

Images and exposure maps were produced for the PN and MOS cameras for both observations in the energy range $0.3-5\keV$ to maximise signal-to-noise. The exposure maps for each instrument were ``spectrally weighted'' by creating several exposure maps produced in narrow energy bands, in which the energy dependence of the vignetting function was small. These maps were weighted by the photon flux predicted in each energy band by the best-fitting spectral model to the global cluster emission (see \textsection \ref{sect:spectra}) and summed. The images and weighted exposure maps for each camera were then combined to give a mosaiced image and map. The exposure map was normalised so that its value at the cluster centroid was 1, and the image was divided by the result. This removes the effects of vignetting and chip gaps, while maintaining as far as possible the Poissonian statistics of the data. An exposure-corrected image of the field of view, combining the data from both observations is shown in Fig \ref{fig:fov}. The diffuse emission from ClJ0046.3$+$8530 is visible near the Northern edge of the field of view, and the increasing PSF distortion with off-axis angle is apparent.

\begin{figure}
\begin{center}
\scalebox{0.4}{\includegraphics*{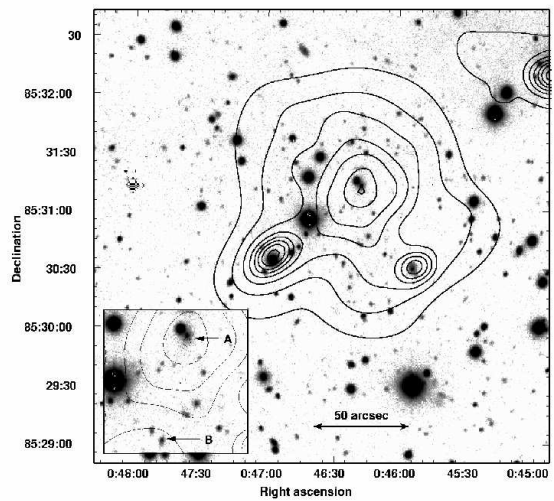}} \\
\caption{\label{fig:xmmoverlay}Contours of X-ray emission detected by \XMM ($0.3-5\keV$) overlaid on combined University of Hawaii $2.2\m$ and WHT R-band images of ClJ0046.3$+$8530. The contours are logarithmically spaced above the lowest contour at $0.32\ctpp$, which is $1.5$ times the background level. The embedded panel is an enlargement of a central $1\arcm\times1\arcm$ region, showing the positions of the galaxies with measured redshifts}
\end{center}
\end{figure}

The exposure-corrected image was adaptively smoothed so that all remaining features were significant at $\ge99\%$ confidence, and contours of the smoothed emission were overlaid on an R-band optical image, as shown in Fig. \ref{fig:xmmoverlay}. Ignoring the distortion of the contours caused by the two point sources to the South, the contours appear fairly circular, indicating that the cluster's atmosphere is reasonably relaxed.

The two galaxies with measured redshifts are labelled in Fig. \ref{fig:xmmoverlay}, and their positions and redshifts are given in Table \ref{tab:pos}. The table also gives the position of the X-ray centroid and the measured X-ray redshift of the cluster (see \textsection \ref{sect:spectra}). All of the following assumes $z=0.624$, the redshift of the central galaxy, as the cluster redshift. 

\begin{table*} 
\begin{tabular}{|c|c|c|c|} \hline 
Object & $\alpha[2000.0]$ & $\delta[2000.0]$ & $z$ \\ \hline
Galaxy A & $00^{\rm h}46^{\rm m}17.73^{\rm s}$ & $+85^{\circ}31\arcm 13.6\arcs$ & $0.6243\pm0.0004$ \\
Galaxy B & $00^{\rm h}46^{\rm m}25.92^{\rm s}$ & $+85^{\circ}30\arcm 34.1\arcs$ & $0.630\pm0.002$ \\
ClJ0046.3$+$8530 & $00^{\rm h}46^{\rm m}17.83^{\rm s}$ & $+85^{\circ}31\arcm 09.5\arcs$ & $0.615^{+0.008}_{-0.006}$ \\ \hline
\end{tabular}
\caption{\label{tab:pos}Positions and redshifts of the two galaxies marked in Fig. \ref{fig:xmmoverlay} and of the cluster itself. The position given for  the cluster is the X-ray centroid, and the redshift is the X-ray redshift (derived in \textsection \ref{sect:spectra}).}
\end{table*} 

\subsection{An extended source to the South East}
\begin{figure}
\begin{center}
\scalebox{0.4}{\includegraphics*{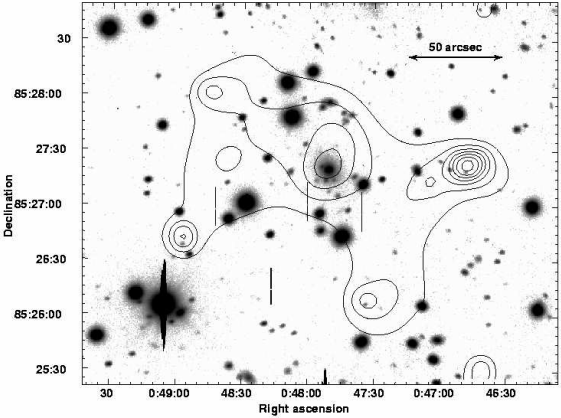}} \\
\caption{\label{fig:large}Contours of a smoothed X-ray image of XMMU J004751.7$+$852722 from both observations and all EPIC detectors overlaid on a WHT R-band image. The contours are logarithmically spaced above the lowest contour at $0.25\ctpp$, which is $1.5$ times the local background level.}
\end{center}
\end{figure}

A region of extended emission (designated XMMU J004751.7$+$852722) was
detected $4\arcm$ South East of ClJ0046.3$+$8530 (see Fig. \ref{fig:fov}). Contours of this
emission (smoothed to $99\%$ significance) are overlaid on an optical
image in Fig. \ref{fig:large}. The centroid of this extended emission
coincides with the 2MASS galaxy 2MASX J00475079$+$8527213
($K=13.5\pm0.2\mags$). Within a radius of $32\arcs$ from the source
centroid, a total of $\approx500$ net photons were detected, using a
local background annulus, and excluding point sources and emission
from ClJ0046.3$+$8530. The properties of XMMU J004751.7$+$852722 are
discussed in \textsection \ref{sect:colmag} and \textsection \ref{sect:xmmu}.

\subsection{Modelling the surface-brightness distribution of ClJ0046.3$+$8530}
The surface-brightness distribution of ClJ0046.3$+$8530 was modelled in \emph{Sherpa} with a two-dimensional (2D) method. A $0.3-5\keV$ image from each detector (without exposure correction) was fit with a surface brightness model which was multiplied by an appropriate exposure map, and convolved with a PSF. Due to the low numbers of counts in a typical pixel, the Cash statistic \citep{cas79} was used when fitting the models. The cluster emission model used was a 2D parameterisation of the $\beta$-model \citep{cav76}. Two additive background model components were used. The first was a simple constant-background model which, after multiplication by the exposure map should accurately represent a flat, but vignetted X-ray background. The second background component consisted of a $0.3-5\keV$ image produced from closed-filter data, to account for the particle background. This particle image was divided by the exposure map, so that after multiplication by the exposure map in the fitting process, the background model consisted of a vignetted and a non-vignetted component (see \citet{mau04a} for a discussion of the background components and these models). 

Due to the large wings of the PSF at the detector position of the cluster, it was not possible to fully exclude point source emission without excluding almost all of the cluster emission. Instead, the point sources within the fitting region were modelled as delta functions, and their position and amplitude were fit to the data. This procedure is not ideal, because the fitting software only allows the model being fit to be convolved with one PSF per dataset, while the shape of the PSF varies over the detector region spanned by the cluster emission (as can be seen in Fig. \ref{fig:xmmoverlay}). \emph{Calview} was used to generate PSF images for the detector position of the cluster centroid at an energy of $1.5\keV$ for each detector (the PSF is not strongly energy dependent in the $0.3-5\keV$ band). The `MEDIUM' accuracy level was used, which provides the only available 2D model of the \XMM PSF, including its arc-like distortion at the cluster off-axis angle.

These models were fit simultaneously to the PN and MOS images from each observation. The results of the fits were not consistent between the observations. The best-fitting $\beta$-profile core radius ($r_c$) and slope ($\beta$) are $r_c=11.5^{+2.8}_{-2.3}\arcs$ and $\beta=0.49^{+0.06}_{-0.02}$ in observation 1, and $r_c=20.2^{+4.4}_{-3.6}\arcs$ and $\beta=0.60^{+0.08}_{-0.03}$ in observation 2. This discrepancy was investigated in detail, and is likely to be caused by a combination of the large PSF, uncertainties in its modelling, and a variation in the flux of brightest point source between the observations. The count rate from the bright point source to the South East was $25\%$ higher in the first observation. Given the uncertainties in modelling the PSF, it is likely that there is more point source contamination remaining in observation 1, increasing the number of counts at large radii, which would cause the $\beta$-profile slope to fit to a lower value. In addition, the core radii found in both observations are small compared to the PSF, which means they may not be reliably measured. This would also affect the values of $\beta$ as there is degeneracy between the two parameters. These fits were repeated with the core radii fixed at $22\arcs$, which corresponds to $150\kpc$, a value typical of local clusters \citep[\egc][]{jon99}. With this constraint, the best-fitting values of all other parameters are consistent between the observations. 

\begin{figure}
\begin{center}
\scalebox{0.6}{\includegraphics*{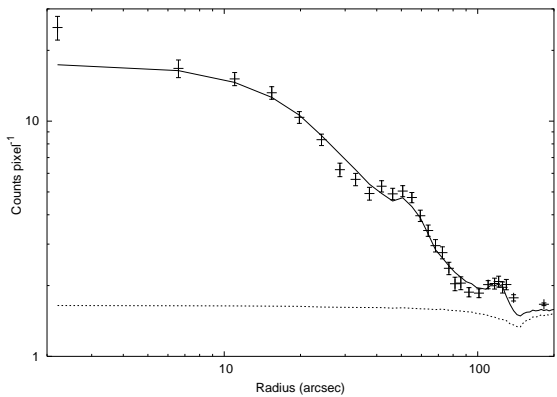}} \\
\caption{\label{fig:2dprof}A radial profile of the best-fitting 2D model to the ClJ0046.3$+$8530 observation 2 data, including instrumental effects, is shown with the solid line. The points are a radial profile of the data, and the dotted line is a profile of the best-fitting background model components.}
\end{center}
\end{figure}

With this in mind, we chose to use only observation 2 in our imaging analysis because it was the longer observation, with better signal-to-noise, and the point source contamination is lower. The best-fitting model has $r_c=20.2^{+4.4}_{-3.6}\arcs$ and $\beta=0.60^{+0.08}_{-0.03}$, with a low ellipticity of $0.07$, and we proceed with these values for all further analysis. With $r_c$ fixed at this value, fitting to the observation 1 data gives $\beta=0.57\pm0.05$, consistent with the above result. Fig. \ref{fig:2dprof} shows a radial profile of the best-fitting 2D model, including instrumental effects, plotted on a profile of the data. The point sources were not excluded from the profile of the data in this comparison, because they are included in the 2D model. The dotted line is a profile of the best-fitting background model components. These profiles show that the model is a reasonably good description of the data, although it is clear again that the point sources at $\approx50\arcs$ and $\approx100\arcs$ are not perfectly modelled. The dip in the profiles at $\approx130\arcs$ is due to the edge of the PN CCD. This is apparent in both the data and the models because the models have been multiplied by exposure maps to replicate the instrumental effects in the data.

In order to measure the extent of the detection, a one-dimensional radial profile of the X-ray emission was extracted from the combined (both observations, all cameras) exposure corrected image. Point sources were excluded out to a radius of $20\arcs$, and the profile was adaptively binned so that all radial bins were significant at the $3\sigma$ level above the background. It is not appropriate to fit a one-dimensional $\beta$-model to this profile, because this would not account for the large PSF. Emission is detected at the $3\sigma$ level out to a radius of $142\arcs$ ($965\kpc$), which corresponds to $88\%$ of $\rt$, as discussed below (\textsection \ref{sect:prop}). While some point source contamination will remain after their exclusion (due to the large PSF wings), the diffuse emission is clearly extended to $\approx140\arcs$ at azimuths with no point sources.

\subsection{Hardness-Ratio Map}
The temperature structure of the cluster was probed with a hardness-ratio map. The hardness ratio, HR, was defined as
\begin{eqnarray}
HR & = & \frac{H-AH_{bg}}{S-AS_{bg}},
\end{eqnarray}
where $H$ and $S$ are the counts in the source region in the hard and soft band respectively, and the $bg$ subscript indicates the counts found in a background region. $A$ is the ratio of the area of the source region to the background region. The error on the hardness ratio is then given by 
\begin{eqnarray}
\sigma(HR)^2 & = & \frac{H+A^2H_{bg}}{(S-AS_{bg})^2} + \frac{(H-AH_{bg})^2(S+A^2S_{bg})}{(S-AS_{bg})^4}.
\end{eqnarray}

\begin{figure}
\begin{center}
\scalebox{0.6}{\includegraphics*{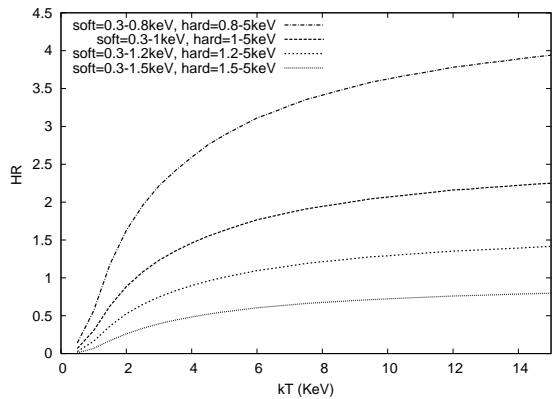}} \\
\caption{\label{fig:hr_t}\MEKAL temperature with corresponding hardness ratio values for different hard and soft energy bands, derived from simulated EPIC spectra. The Galactic absorption, redshift and instrumental responses appropriate to ClJ0046.3$+$8530 were used.}
\end{center}
\end{figure}

Before defining the hard and soft bands, absorbed \MEKAL \citep{kaa93} spectra were simulated for the three EPIC detectors (with the appropriate redshift, Galactic absorption and instrumental response) to estimate the sensitivity of HR to gas temperature variations, with different hard and soft energy bands. A soft band of $0.3-1\keV$, and a hard band of $1-5\keV$ were chosen because HR values in the range $0.5-2$ (within which the statistics are not dominated by one band with low counts) corresponded to a temperature range of $\approx1-10\keV$, a reasonable range about the global temperature. Fig. \ref{fig:hr_t} illustrates this, showing the relationship between hardness ratio and \MEKAL temperature for different hard and soft bands.

\begin{figure}
\begin{center}
\scalebox{0.4}{\includegraphics*{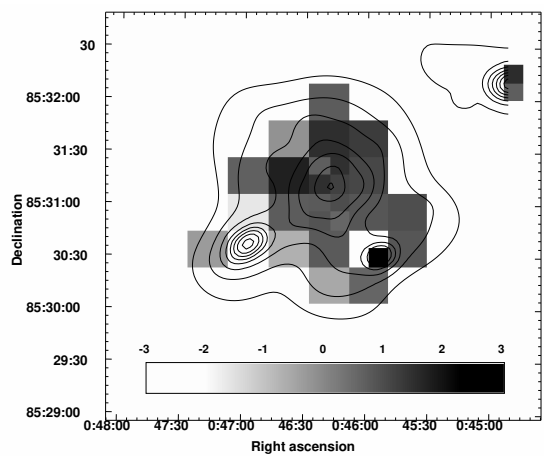}} \\
\caption{\label{fig:HStsig}Adaptively binned HR significance map of ClJ0046.3$+$8530 data. The overlaid contours are the same as in Fig. \ref{fig:xmmoverlay} and the scale-bar shows the HR significance colour scaling.}
\end{center}
\end{figure}

An adaptive binning template was computed from a broad-band image, to define image bins with $>100$ counts in order to minimise statistical uncertainties while keeping reasonable spatial resolution. A few bins were allowed to fall below this threshold to improve the resolution, but all bins with $<50$ counts were excluded from further analysis. The background-subtracted, exposure-corrected hard- and soft-band images were binned according to this template, and hardness ratios with errors were computed for each bin. Fig. \ref{fig:HStsig} shows a significance map where the value in each bin is given by the difference in HR between the data at that point, and the HR corresponding to the global temperature, divided by the error on the HR in that bin. 

All of the bins (excluding those associated with the point sources) are consistent within the statistical uncertainties with the emission being isothermal; $17$ of $25$ ($68\%$) are within $1\sigma$ of the global HR and all bins are within $2\sigma$. However, as the larger image bins in the HR significance map are similar in size to the $75\%$ encircled energy radius of the PSF, this method would only be sensitive to reasonably strong temperature variations. Of the two point sources, that to the South West ($\alpha[2000.0]=00^{\rm h}45^{\rm m}53.81^{\rm s}$, $\delta[2000.0]=+85^{\circ}30\arcm 30.6\arcs$) has harder emission (at the $2.7\sigma$ level), which may be consistent with it being an absorbed AGN in the foreground galaxy that appears to be the counterpart of the X-ray point source in Fig. \ref{fig:xmmoverlay}. The South East point source ($\alpha[2000.0]=00^{\rm h}46^{\rm m}57.62^{\rm s}$, $\delta[2000.0]=+85^{\circ}30\arcm 36.6\arcs$) has significantly ($\approx10\sigma$) softer emission, and appears to be associated with a foreground star. The spectra of these sources and the galaxy cluster are discussed in \textsection \ref{sect:spectra}.

\section{X-ray spectral analysis}\label{sect:spectra}
Spectra were extracted for all EPIC detectors for both observations. For comparison, background spectra were extracted both locally, and from blank-sky background files. The results of the spectral analysis were found to be independent of the background type used, so local background spectra were used in all further analysis, as the systematic uncertainties involved were considered to be smaller. The source spectra were extracted within a circle of radius $88\arcs$ centred on the cluster centroid, corresponding approximately to the $3\sigma$ detection radius in each separate observation (as opposed to the $3\sigma$ detection radius of $142\arcs$ when the observations were combined). The strong radial dependence of effective area at the cluster position (Fig. \ref{fig:effA}) was taken into consideration when selecting the regions from which to extract the local background spectra. Exposure maps of each detector were used to select spatial regions where the mean effective exposure (equivalent to effective area) was close to that in the source spectrum extraction region. Ancillary response files (ARFs), weighted by the spatial distribution of counts in the source region, were also generated for the cluster's off-axis position. 

The two point sources to the South fell within the source region, and their emission was excluded to a radius of $20\arcs$ ($\approx75\%$ encircled energy radius of the PSF of all \XMM telescopes at $0.3-5\keV$). Due to the large PSF wings, and its elongation at this off-axis angle, some point source contamination will remain, and the effect of this is examined below.

\begin{figure}
\begin{center}
\scalebox{0.3}{\includegraphics*[angle=270]{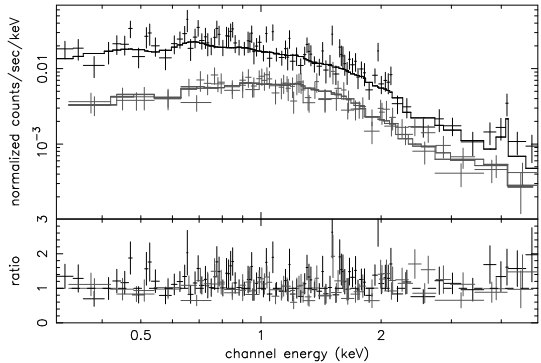}} \\
\caption{\label{fig:spec}All EPIC spectra for observations 1 and 2 of ClJ0046.3$+$8530, and the best-fitting absorbed \MEKAL model. Absorption is frozen at the Galactic value. For presentation purposes, the data from the same instruments in the two observations have been combined for this plot, and adjacent bins have been combined to give a minimum signal-to-noise ratio of 3. The lower panel shows the ratio of data to model values.}
\end{center}
\end{figure}

The spectra from the three EPIC detectors in the $0.3-5\keV$ band were each grouped to give $\ge20$ counts per energy bin, and were fit simultaneously with an absorbed \MEKAL model. This was done for the two observations independently, and the best-fitting model parameters were consistent. All spectra for both observations were then fit simultaneously with an absorbed \MEKAL model. The best-fitting temperature and metallicity are $kT=4.1\pm0.3\keV$ and $Z=0.39^{+0.14}_{-0.13}$ respectively, when the absorbing column was frozen at the Galactic value of $7.6\times10^{20}\pcmsq$ \citep{dic90}, with $\chisq/\nu=372/360$. This model is shown with the data in Fig. \ref{fig:spec}. When the absorbing column is allowed to vary, the best-fitting value is $(6.8\pm1.2)\times10^{20}\pcmsq$, consistent with the Galactic value. In addition, an X-ray redshift was measured, and the best-fitting value is $z=0.615^{+0.008}_{-0.006}$, which supports the optical redshift of $0.624$ ascribed to the cluster based on the two measured galaxy redshifts.

The spectrum of the soft, bright point source to the South East was examined, using a local annulus as a background region, to ensure that the cluster emission is also included in the background spectrum. The spectrum is well fit by an unabsorbed \MEKAL model ($\chisq/\nu=9.95/10$) of $kT=0.53^{+0.13}_{-0.10}\keV$ and $Z=0.06\pm0.03\Zsol$. The redshift of this model was frozen at zero. A power-law model is rejected at the $95\%$ level by the data, supporting the conclusion that the emission is stellar. The unabsorbed flux from this point source is $(8.4\pm1.9)\times10^{-15}\flux$ in the $0.5-2\keV$ band.

To test whether the emission from this star affected the cluster spectrum, a cluster spectrum was extracted including this source, and its emission was included in the model. The model used was an absorbed \MEKAL model at $z=0.624$ plus an unabsorbed \MEKAL model at $z=0$. The best-fitting temperatures of the cluster and star are $kT=4.8^{+1.2}_{-0.8}\keV$ and $0.38^{+0.18}_{-0.08}\keV$ respectively, with $\chisq/\nu=112.4/121$. The good agreement between these fits, and those found when fitting the sources independently indicates that the cluster emission is not seriously contaminated by the star, and most importantly, that the temperature measured for the cluster is not influenced by any remaining stellar emission. 

A cluster spectrum was also extracted with the emission from the harder South-West point source included. When this spectrum was fit with an absorbed \MEKAL model, the best-fitting temperature was $4.9\pm0.5\keV$, hotter than the best-fitting temperature when that point source was excluded. This is consistent with the harder spectrum of that source. When an additional absorbed power-law component with a spectral index of $1.5$ was included in the model, the best-fitting \MEKAL temperature reduced to $3.8\pm1.0\keV$, which agrees well with the cluster temperature measured when the South-West point source was excluded. This suggests that the cluster spectrum was not significantly contaminated by South-West point source after it was excluded.

As a further test, the cluster emission model was fit to the data with the point sources excluded, in the energy range $2-8\keV$. In this band the contribution from the soft stellar source should be negligible. The best-fitting temperature is $kT=4.1^{+1.1}_{-0.8}\keV$, providing further evidence that the measured temperature was not affected by point source contamination.

The flux of ClJ0046.3$+$8530 measured by \ROSAT in the $0.5-2\keV$ passband is $1.6\times10^{-13}\flux$ \citep{hor04} when extrapolated to an infinite radius. For comparison, the \XMM flux in this band is $(1.45\pm0.07)\times10^{-13}\flux$ (after extrapolation to infinite radius). This difference in flux is due to the two point sources, which were unresolved (or barely resolved in one case) in the \ROSAT data, and included in the \citet{hor04} flux measurement.

\subsection{Temperature Profile}\label{sect:tprof}
\begin{figure}
\begin{center}
\scalebox{0.6}{\includegraphics*{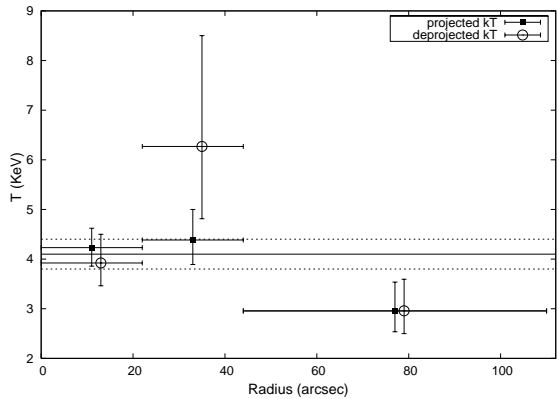}} \\
\caption{\label{fig:tprof}Projected and deprojected temperature profiles of ClJ0046.3$+$8530, based on spectra fit with abundance frozen at $0.3\Zsol$, and Galactic absorption. The solid line is the best-fitting global (emission weighted) temperature, with $1\sigma$ errors represented by the dashed lines. The temperature error bars have been offset by $2\arcs$ on the deprojected data points for clarity.}
\end{center}
\end{figure}

With $\sim5000$ net counts (all EPIC detectors combined) within the detection radius, it was possible to measure a temperature profile of the gas in ClJ0046.3$+$8530. Spectra were extracted in 3 annular bins for all detectors, and both observations, and fit simultaneously with an absorbed \MEKAL model, with the abundance frozen at $0.3\Zsol$, and the absorbing column frozen at the Galactic value. A locally extracted background spectrum was used. The best-fitting temperatures are shown in Fig. \ref{fig:tprof}, and are consistent (with large errors) with the global temperature of $4.1\pm0.3\keV$, though the outer bin could indicate cooler gas at larger radii. The spectral fit to the data from the outer bin was also performed in the $2-8\keV$ band, to test the effect of any contaminating emission from the bright point source. The best-fitting temperature agrees well (with large errors) with that shown in Fig. \ref{fig:tprof}. The data do not suggest any cool gas in the central regions.

The effect of emission from gas in the outer annuli being projected onto the inner annuli
was modelled with an `onion skin' method. The temperature structure was modelled as a series of spherical shells (each of which was isothermal), and the spectra were fit from the outermost shell in. The spectrum of a shell was modelled with a single temperature \MEKAL component, plus a \MEKAL component for each external shell, whose temperature was fixed at the value measured in that shell, and whose normalisation was multiplied by a scale factor. These scale factors accounted for the volume of each external shell along the line of sight to the shell being fit, and the variation in density across each external shell using the measured gas density profile. The deprojected temperatures are also plotted in Fig. \ref{fig:tprof}. The form of this deprojected profile is not significantly different from the projected profile, and does not reveal any central cool gas, although the size of the errors is increased, as one would expect, as there were less photons available to constrain the temperature of the free component in the interior bins.

It should be noted that the effect of the large PSF mixing the emission between annular bins was not taken into account here. The $90\%$ encircled energy radius of the PSF at the position of the cluster at $1.5\keV$ is $\approx65\arcs$ for the PN detector. This means that it is possible there was a temperature gradient that was blurred out by the PSF. We derived the total gravitating mass below assuming isothermality, which is consistent with the observed deprojected temperature profile, but we also discuss the possible effects of a temperature gradient.

\begin{figure}
\begin{center}
\scalebox{0.6}{\includegraphics*{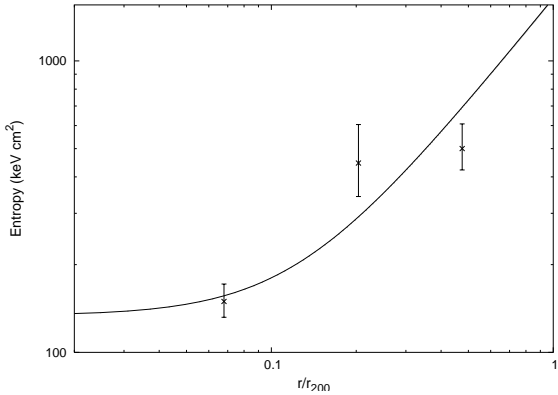}} \\
\caption{\label{fig:eprof}Entropy of ClJ0046.3$+$8530 is plotted against radius as a fraction of $\rt$. The solid line shows the entropy derived assuming isothermality at the global temperature, and the data points correspond to the deprojected temperature bins in Fig. \ref{fig:tprof}}
\end{center}
\end{figure}

This temperature profile enabled the measurement of the entropy profile of the cluster gas, where entropy ($S$) is defined here as $S=T/n_e^{2/3}\ent$. The electron density profile was derived from the best-fitting surface-brightness profile, normalised to the central value derived from the measured \MEKAL normalisation (see \citet{mau03a}), and assuming spherical symmetry. The entropy profile is shown in Fig. \ref{fig:eprof}. The entropy at $0.1\rt$ (with $\rt$ defined as described in \textsection \ref{sect:prop}) is $180\pm24\ent$, with the error taken from the fractional temperature error on the nearest bin in the deprojected profile. This is lower than the value of $\approx300\ent$ found in local systems of the same global temperature by \citet{pon03}. This difference in entropy is compared with the predicted self-similar evolution in \textsection \ref{sect:dis}.

\section{Determination of Global Properties}\label{sect:prop}
The properties of ClJ0046.3$+$8530 were derived following the methods described in \citet{mau04a}. In summary, the total mass profile was derived assuming hydrostatic equilibrium, spherical symmetry, and isothermality, and used to define an overdensity profile with respect to the critical density at the redshift of observation. The virial radius, $\rt$, was then defined as the radius corresponding to an overdensity of $200$. The central gas density was computed from the measured \MEKAL normalisation, and used to normalise the gas density profile. The errors on all the non-measured quantities were derived by $10,000$ Monte-Carlo randomisations based on the errors on the measured properties.

Our assumption of isothermality is supported by the measured temperature profile and hardness ratio map, while the circular appearance of the X-ray contours and low measured ellipticity support the assumptions of hydrostatic equilibrium and spherical symmetry.

The derived mass profile of ClJ0046.3$+$8530 enabled the definition of $\rt=1.10^{+0.06}_{-0.07}\Mpc$, which means that the cluster emission is detected out to $0.88\rt$ and the temperature profile is measured out to $0.68\rt$. This greatly reduces the uncertainties in extrapolating the cluster properties to $\rt$. The gas mass fraction of the system at $\rt$ is found to be $16\pm4\%$. The properties of ClJ0046.3$+$8530 are summarised in Table \ref{tab:summary}.

\begin{table*} 
\scalebox{0.8}{
\begin{tabular}{|c|c|c|c|c|c|c|c|c|c|} \hline 
Cluster & Redshift & $T (keV)$ & $L_{bol} (erg\s^{-1})$ & $r_c ({\kpc})$ & $\beta$ & $\rt (Mpc)$ & $M_{gas}(\rt) (M_\odot)$ & $M_{tot}(\rt) (M_\odot)$ \\ \hline
ClJ0046.3$+$8530 & $0.624$ & $4.1\pm0.3$ & $(4.3\pm0.3)\times10^{44}$ & $137^{+30}_{-24}$ & $0.60^{+0.08}_{-0.03}$ & $1.10^{+0.06}_{-0.07}$ & $(4.7\pm0.6)\times10^{13}$ & $3.0^{+0.6}_{-0.5}\times10^{14}$ \\ \hline
\end{tabular}
}
\caption{\label{tab:summary}Summary of the measured and inferred properties of cluster ClJ0046.3$+$8530 based on \XMM observations, assuming a cosmology of $\Omega_{M}=0.3$ $(\Omega_\Lambda=0.7)$ and $H_0=70$\kmpspMpc. The luminosity and masses were scaled out to $\rt$.}
\end{table*}

\section{Optical imaging analysis}\label{sect:colmag}
\begin{figure}
\begin{center}
\scalebox{0.6}{\includegraphics*{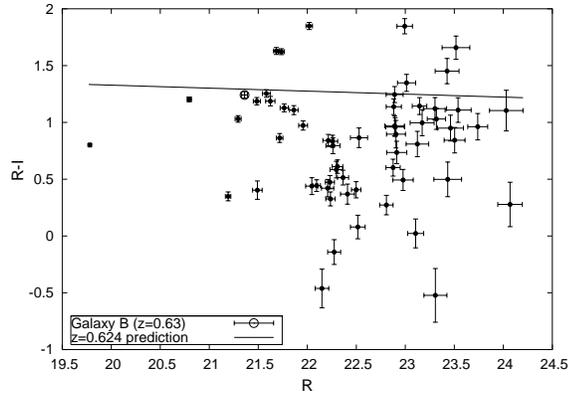}} \\
\caption{\label{fig:colmag_j0046}Colour-magnitude diagram for ClJ0046.3$+$8530 using galaxies within $60\arcs$ of the X-ray centroid. Galaxy B in ClJ0046.3$+$8530 with a spectrally confirmed redshift is marked as a solid square. The line shows the predicted red sequence at $z=0.624$ (see text).}
\end{center}
\end{figure}

\begin{figure}
\begin{center}
\scalebox{0.6}{\includegraphics*{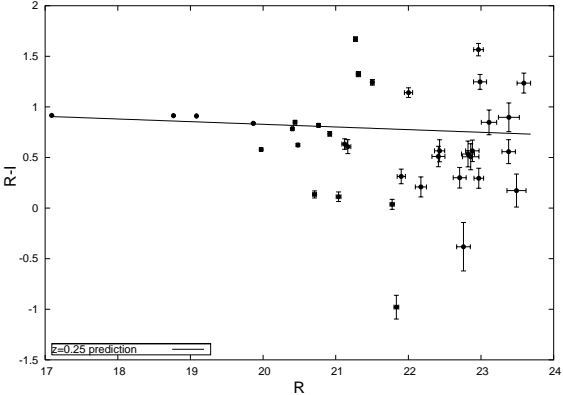}} \\
\caption{\label{fig:colmag_j0047}Colour-magnitude diagram for XMMU J004751.7$+$852722 using galaxies within $65\arcs$ of the X-ray centroid. The line shows the predicted red sequence at $z=0.25$ (see text).}
\end{center}
\end{figure}

The R and I band WHT images were used to investigate the properties of
the galaxies in ClJ0046.3$+$8530 and XMMU J004751.7$+$852722. Galaxies
were selected by using SExtractor \citep{ber96}, and their $R-I$
colours and total magnitudes were measured. Colours were measured
using apertures chosen to maximise signal to noise, and total
magnitudes were obtained from the SExtractor MAG\_AUTO
measurements. Magnitudes and colours were corrected for Galactic extinction using the measurements of \citet{sch98}. The galaxies within a radius $65\arcs$ of the X-ray
centroid of the XMMU J004751.7$+$852722 and within $60\arcs$
($0.4\rt$) of the X-ray centroid
of ClJ0046.3$+$8530 were then selected. These are plotted on a
colour-magnitude diagram in Figs. \ref{fig:colmag_j0046} and
\ref{fig:colmag_j0047}. Galaxy B in ClJ0046.3$+$8530, with a spectrally
confirmed redshift, is marked separately. The brightest cluster galaxy
(galaxy A in Fig. \ref{fig:xmmoverlay}) was partially blended with a
bright star, leading to large uncertainties on its colour and
magnitude, so is not shown here. There is a possible red sequence at
$R-I\approx1.3$ in the  ClJ0046.3$+$8530 galaxies, while a
stronger sequence at $R-I\approx1$ is clear in the galaxies of XMMU
J004751.7$+$852722. This indicates that XMMU J004751.7$+$852722 is at
a lower redshift than ClJ0046.3$+$8530, perhaps unsurprising given the 
coincidence of the source with the relatively bright 2MASS galaxy. 

To estimate the redshift of XMMU J004751.7$+$852722, the stellar 
 population synthesis models of \citet{bru93} were used to
predict the normalisation of the red sequence vs. redshift.
A single stellar population was used
with a burst of star formation at z$_f$=3 and passive evolution
 thereafter, a Salpeter IMF
and solar metallicity. These are typical values found for
cluster elliptical evolution studies \citep[\egc][]{ell04}.
Similar results are obtained if z$_f$=2 is
 assumed. The predicted colours were normalised to those of the Coma cluster at
 $L*$ \citet{eis04}.
The resulting predictions have been plotted in Figs. \ref{fig:colmag_j0046} and
\ref{fig:colmag_j0047}
for redshifts z=0.624 and z=0.25, using appropriate 
values for the slope from the Coma cluster compilation of
\citet{eis04}. The prediction matches the observed colours
of the red sequence galaxies in ClJ0046.3$+$8530, and a redshift of
z=0.25 is a good  match to the red sequence in XMMU J004751.7$+$852722.

The BCG in XMMU J004751.7$+$852722
has R=17.4 and K$\approx$13.5, at the limit of the 2MASS
sensitivity. The R-K colour is consistent with that of local elliptical
galaxies when the k-corrections are taken into account. Assuming the redshift 
of $z=0.25$ is correct, the
absolute magnitude is $M_R\approx -23.4$+5log($h_{70}$), and this is 
a luminous giant elliptical galaxy, comparable in luminosity 
to BCGs in some rich, high $L_X$ systems.


\section{Properties of the nearby cluster  XMMU J004751.7$+$852722}\label{sect:xmmu}
A hardness ratio was computed for XMMU J004751.7$+$852722, using the
energy bands defined above, giving $HR=1.74\pm0.24$. This HR value was
converted to a temperature using simulated spectra as described above,
with Galactic absorption and a metallicity of $0.3\Zsol$. As the
redshift of the source was not certain, spectra were simulated at a
range of redshifts, giving a range of possible temperatures. 
At $z=0$ the HR value corresponds to $kT=3.0^{+2.0}_{-1.1}\keV$,
increasing to $kT=4.8^{+2.3}_{-1.4}\keV$ at the estimated redshift
$z=0.25$. The measured count rate was then converted to a $0.5-2\keV$
unabsorbed flux, using these derived temperatures, giving
$F_X=5.4\times10^{-15}\flux$ ($kT=3.0\keV$) and
$F_X=6.2\times10^{-15}\flux$ ($kT=4.8\keV$). 
The corresponding bolometric X-ray luminosity at z=0.25 is $L_X\approx3\times10^{42}\ergps$ (within the
measurement radius).
We conclude that it is likely that this extended source is a cluster
at z$\approx$0.25 with low $L_X$.

\section{Discussion}\label{sect:dis}
The evolution of cluster entropy was investigated by comparing the gas entropy measured at $0.1\rt$ in ClJ0046.3$+$8530 with that measured in local systems by \citet{pon03}. Under self-similar models, entropy is simply proportional to temperature. However with a large sample of systems with high-quality data, \citet{pon03} found that this was not the case. To remove the effect of temperature dependence, ClJ0046.3$+$8530 was compared with a data bin of eight systems in the temperature range $3.6-4.6\keV$ with $\bar{kT}=4.1\pm0.4\keV$ ($\bar{z}=0.036$). The entropy at $0.1\rt$ in ClJ0046.3$+$8530 is $S=180\pm24\cmsq$ and the mean entropy of the eight local systems at the same radius is $S=327\pm42\cmsq$.

The mean density within a given overdensity radius (relative to the critical density) is proportional to $H(z)^2$, assuming self-similar scaling. Thus, the redshift dependence of scaled entropy is given by $(H_0E(z))^{-4/3}$, where 
\begin{eqnarray}
E(z) & = & (1+z)\left(1+z\Omega_M+\frac{\Omega_\Lambda}{(1+z)^2}-\Omega_\Lambda\right)^{1/2}.
\end{eqnarray}

Using this formulation, the entropy in ClJ0046.3$+$8530 was scaled to the mean redshift of the local systems (assuming that the systems virialised at the redshift at which they are observed), giving $S=285\pm40\cmsq$, which is consistent with the local systems. This result is consistent with a simple self-similar evolution of entropy with redshift, a result also seen in the massive, relaxed cluster ClJ1226.9$+$3332 at $z=0.89$ \citep{mau04a}. This result will be investigated further with a larger sample of high-z clusters in a future paper.

In the context of other high-redshift ($z\ge0.6$) clusters observed with \Chandra and \XMM \citep[\egc][]{vik02,arn02b,hol02,mau03a,mau04a,ros04a}, ClJ0046.3$+$8530 has an average temperature and mass. The long exposure time of these two observations enabled the system to be studied in more detail than is usually possible at high redshift. The measurement of the gas temperature structure and detection of emission to approaching the virial radius is relatively rare; temperature profiles have been measured in only three other $z\ge0.6$ clusters, MS 1054-0321 \citep{jel01}, RX J1120.1$+$4318 \citep{arn02b} and ClJ1226.9$+$3332 \citep{mau04a}.

The total gravitating mass of $3.0^{+0.6}_{-0.5}\times10^{14}\Msol$ derived for the system assumed isothermality, which is consistent with the observed temperature profile, and also supported by the hardness-ratio map. Thus this mass should be reasonably reliable. However, it is also possible that there is a temperature gradient present, which was not detected because of the effect of the telescope PSF. In a sample of 66 local systems with measured temperature profiles \citet{san03} found that the incorrect assumption of isothermality leads to an average overestimate of the total mass by $\approx30\%$ at $\rt$. Thus this is a reasonable estimate of the possible systematic error in our total mass measurement.

The measured metal abundance in ClJ0046.3$+$8530 is consistent with the value of $0.3\Zsol$ found in samples of low- and moderate-redshift clusters \citep[\egc][]{mus97a}. In particular, the data impose a $3\sigma$ lower limit of $Z>0.17\Zsol$, which supports the growing evidence of a high redshift ($z>1$) of enrichment of the ICM \citep[\egc][]{mus97a,toz03}. 

The gas mass fraction of $f_{gas}=0.16\pm0.04$ derived at $\rt$ for ClJ0046.3$+$8530 should be a reasonably reliable measurement, because of the large extent of the surface-brightness and temperature profiles. This value is in good agreement with the mean gas mass fraction of $f_{gas}=0.13\pm0.01$ (also at $\rt$) found by \citet{san03} in their large sample of local systems. In a sample of 6 relaxed clusters in the redshift range $0.1<z<0.5$ observed with \Chandra, and with reliable masses confirmed by gravitational lensing, \citet{all02a} found a mean gas mass fraction of $0.113\pm0.005$. While those measurements were made at a radius of r$_{2500}$ (corresponding to a mean internal density of 2500 times the critical density), the value we derive for ClJ0046.3$+$8530 at $\rt$ is consistent with the \citet{all02a} measurements. 

In addition, the gas density slope parameter ($\beta$) measured here for ClJ0046.3$+$8530 is consistent with those of similar temperature clusters in the \citet{san03} sample. Overall, these results suggest that the enrichment and distribution of the ICM is similar in at least some high-redshift clusters to that in local clusters.

\section{Conclusions}
A detailed \XMM study of the high-redshift galaxy cluster ClJ0046.3$+$8530 (z=0.624) was conducted. The circular appearance of the isophotes, and the low ellipticity of the best-fitting surface brightness model suggest that the ICM is reasonably relaxed. A temperature profile and hardness-ratio map show no significant temperature structure. The metal abundance, gas-density profile slope, and gas-mass fraction of ClJ0046.3$+$8530 are all consistent with those measured in local clusters. The entropy of ClJ0046.3$+$8530 measured at $0.1\rt$ is consistent with a self-similar model, in which high-z clusters are identical to their low-z counterparts, with their properties scaled by the increasing (with redshift) density of the universe in which they formed.

The properties of ClJ0046.3$+$8530, along with 10 other high-redshift clusters from the WARPS, will be used to investigate the evolution of the scaling relations between cluster properties in a forthcoming paper.

\section{Acknowledgements}
We thank Simon Ellis for running the  Bruzual-Charlot stellar 
 population synthesis models to estimate the redshift of XMMU J004751.7$+$852722. We are grateful to Alastair Sanderson for providing access to the local cluster data, and the estimates of the systematic uncertainties introduced by assuming isothermality. BJM is supported by a PPARC postgraduate studentship. HE gratefully acknowledges financial support from NASA grant NAG 5-10085. We acknowledge use of the ING service programme, La Palma. This publication makes use of data products from the Two Micron All Sky Survey, which is a joint project of the University of Massachusetts and the Infrared Processing and Analysis Center/California Institute of Technology, funded by the National Aeronautics and Space Administration and the National Science Foundation.

\bibliographystyle{mn2e}
\bibliography{clusters}

\end{document}